\documentclass[fleqn,usenatbib]{mnras}

\usepackage{newtxtext,newtxmath}
\usepackage{hyperref}

\usepackage[T1]{fontenc}

\DeclareRobustCommand{\VAN}[3]{#2}
\let\VANthebibliography\thebibliography
\def\thebibliography{\DeclareRobustCommand{\VAN}[3]{##3}\VANthebibliography}

\def\mearth{{\rm\,M_\oplus}}

\usepackage{graphicx}	% Including figure files
\usepackage{amsmath}	% Advanced maths commands
\title[Dynamical stability of perturbed rings of co-orbital planets]{Survival and dynamics of rings of co-orbital planets under perturbations}

\author[Raymond, Veras, Clement, Izidoro, Kipping \& Meadows]{Sean N. Raymond$^1$\thanks{E-mail: rayray.sean@gmail.com}, 
Dimitri Veras$^\mathrm{2,3,4}$, 
Matthew S. Clement$^{5,6}$, 
Andre Izidoro$^{7,8}$,
\newauthor David Kipping$^9$ and 
Victoria Meadows$^{10,11}$
\\
$^\mathrm{1}$Laboratoire d'Astrophysique de Bordeaux, CNRS and Universit{\'e} de Bordeaux, All{\'e}e Geoffroy St. Hilaire, 33165 Pessac, France \\
$^2$Centre for Exoplanets and Habitability, University of Warwick, Coventry CV4 7AL, UK\\
$^3$Centre for Space Domain Awareness, University of Warwick, Coventry CV4 7AL, UK\\
$^4$Department of Physics, University of Warwick, Coventry CV4 7AL, UK\\
$^5$Earth and Planets Laboratory, Carnegie Institution for Science, 5241 Broad Branch Road, NW, Washington, DC, 20015, USA\\
$^6$Johns Hopkins APL, 11100 Johns Hopkins Rd, Laurel, MD 20723, USA\\
$^7$Department of Physics and Astronomy, 6100 Main MS-550, Rice University, Houston, TX 77005, USA \\
$^8$Department of Earth, Environmental and Planetary Sciences, MS 126, Rice University, Houston, TX 77005, USA\\
$^9$Department of Astronomy, Columbia University, 550 West 120th Street, New York, NY 10027, USA\\
$^{10}$Department of Astronomy and Astrobiology Program, University of Washington, Box 351580, Seattle, WA 98195, USA\\ 
$^{11}$NASA Astrobiology Institute's Virtual Planetary Laboratory, Box 351580, University of Washington, Seattle, WA 98195, USA
}

\date{Accepted XXX. Received YYY; in original form ZZZ}

\pubyear{2022}

\begin{document}
\label{firstpage}
\pagerange{\pageref{firstpage}--\pageref{lastpage}}
\maketitle

\begin{abstract}
In co-orbital planetary systems, two or more planets share the same orbit around their star.  Here we test the dynamical stability of co-orbital rings of planets perturbed by outside forces.  We test two setups: i) 'stationary' rings of planets that, when unperturbed, remain equally-spaced along their orbit; and ii) horseshoe constellation systems, in which planets are continually undergoing horseshoe librations with their immediate neighbors. We show that a single rogue planet crossing the planets' orbit more massive than a few lunar masses ($0.01-0.04 \mearth$) systematically disrupts a co-orbital ring of 6, 9, 18, or 42 Earth-mass planets located at 1 au. Stationary rings are more resistant to perturbations than horseshoe constellations, yet when perturbed they can transform into stable horseshoe constellation systems. Given sufficient time, any co-orbital ring system will be perturbed into either becoming a horseshoe constellation or complete destabilization.
\end{abstract}

\begin{keywords}
planets and satellites: dynamical evolution and stability -- extraterrestrial intelligence -- astrobiology.
\end{keywords}

%%%%%%%%%%%%%%%%%%%%%%%%%%%%%%%%%%%%%%%%%%%%%%%%%%

%%%%%%%%%%%%%%%%% BODY OF PAPER %%%%%%%%%%%%%%%%%%

\section{Introduction}
In a co-orbital system, two or more planets share the same orbit.  The best-known cases of co-orbital systems are Jupiter's Trojan asteroids, whose orbits oscillate ("librate") about Jupiter's L4 and L5 Lagrange points, 60 degrees ahead of and behind its orbit, and Saturn's moons Janus and Epimetheus, which follow horseshoe orbits~\citep{smith80,dermott81b}.  It is possible that co-orbital planetary systems may be common, although they remain to be discovered~\citep[e.g.][]{rowe06}.  Simulations that include gas-driven migration frequently produce co-orbital planets in Trojan configurations, with two planets in each other's mutual L4/L5 points~\citep{cresswell09,izidoro17,raymond18b}, and sometimes in horseshoe configurations as well~\citep{rodriguez19}.  

\cite{salo88} showed that co-orbital rings of planets could remain stable and maintain a fixed orbital spacing.  The conditions for stability were that the ring must contain at least 6 planets, which must have equal masses.  The planets must be evenly-spaced around the star with a separation between planets of at least 12 mutual Hill radii $\mathrm{R_{H,m}}$.  The mutual Hill radius is defined as $\mathrm{R_{H,m}} = \mathrm{a (2 m_p/3 M_\star})^{1/3}$, where $\mathrm{a}$ is the orbital radius, $\mathrm{m_p}$ is the planet mass, and $\mathrm{M_\star}$ is the stellar mass.  For Earth-mass planets at 1 astronomical unit around a Solar-mass star, up to 42 planets can remain on stable orbits for Gyr timescales~\citep{smith10}.  We refer to these as `stationary' rings of planets because their relative positions -- when viewed in a co-rotating frame -- remain fixed.

In a recent paper, we demonstrated the existence of {\em horseshoe constellations}, which represent a different type of co-orbital ring of planets~\citep{raymond23}.  In these systems, co-orbital planets do not remain stationary relative to one another, but rather undergo horseshoe oscillations with their immediate neighbors. A horseshoe constellation system at 1 au can contain at least 24 Earth-mass planets and remain stable for billions of years. 

How easily do co-orbital rings of planets disrupt in the face of perturbations?  This question is of interest, because the discovery of such systems is within the reach of current exoplanet observations~\citep[see, for instance, the TROY project: ][]{lillobox18a,lillobox18b}. In a previous paper~\citep{raymond23} we discussed possible formation pathways of systems containing many co-orbital planets.  While natural pathways may exist (for instance, via fragmentation or coagulation within a ring of material around a young star), the natural formation of a ring of co-orbital planets is a very low-probability event.  Rather, one might imagine that any such system may have instead been engineered by a highly-advanced civilization.\footnote{Indeed, this scientific blog post dedicated to such systems invokes that such systems must be 'Engineered': see \url{https://planetplanet.net/2017/05/03/the-ultimate-engineered-solar-system/}.}  

\begin{figure*}
	\includegraphics[width=\columnwidth]{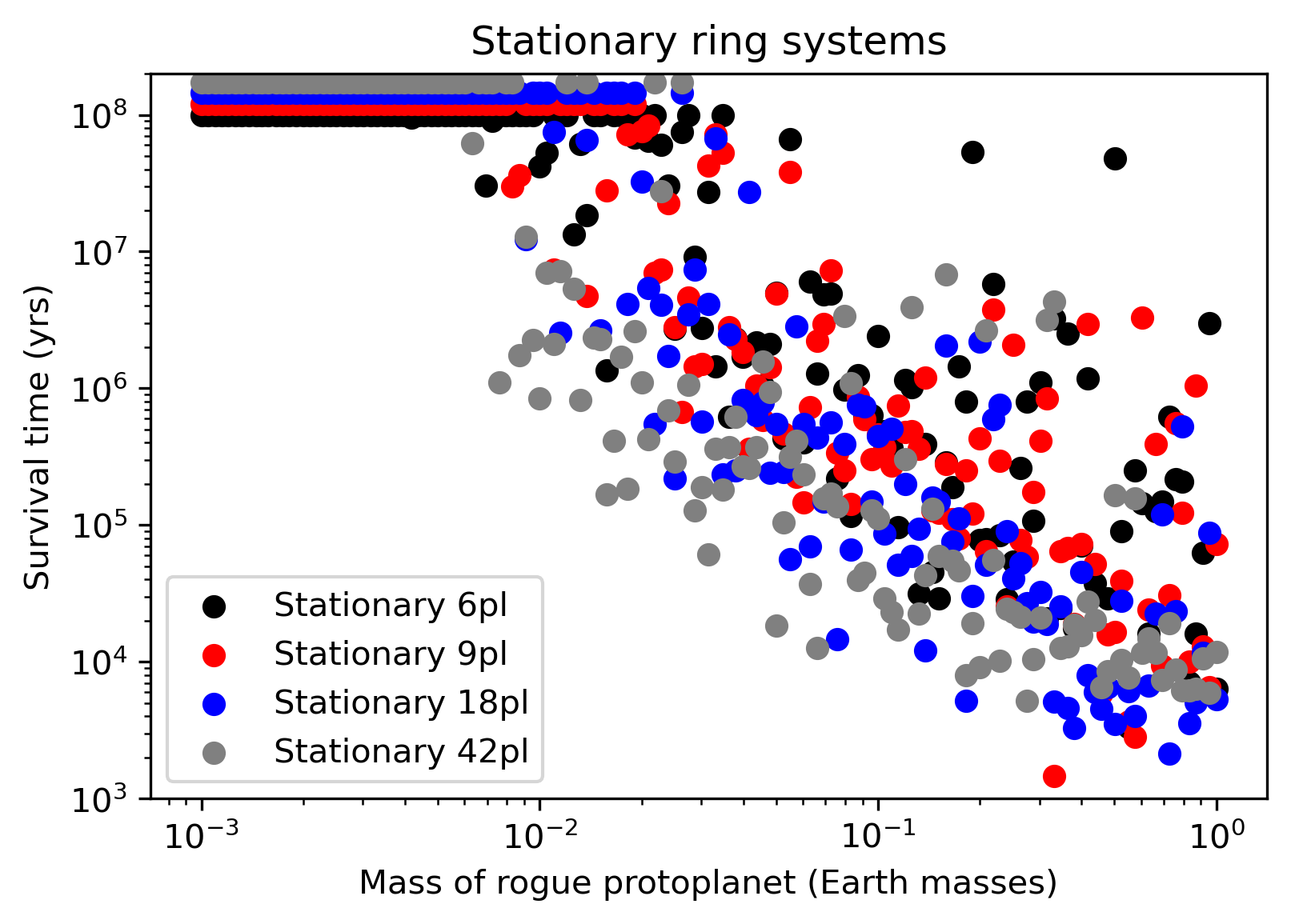}
 	\includegraphics[width=\columnwidth]{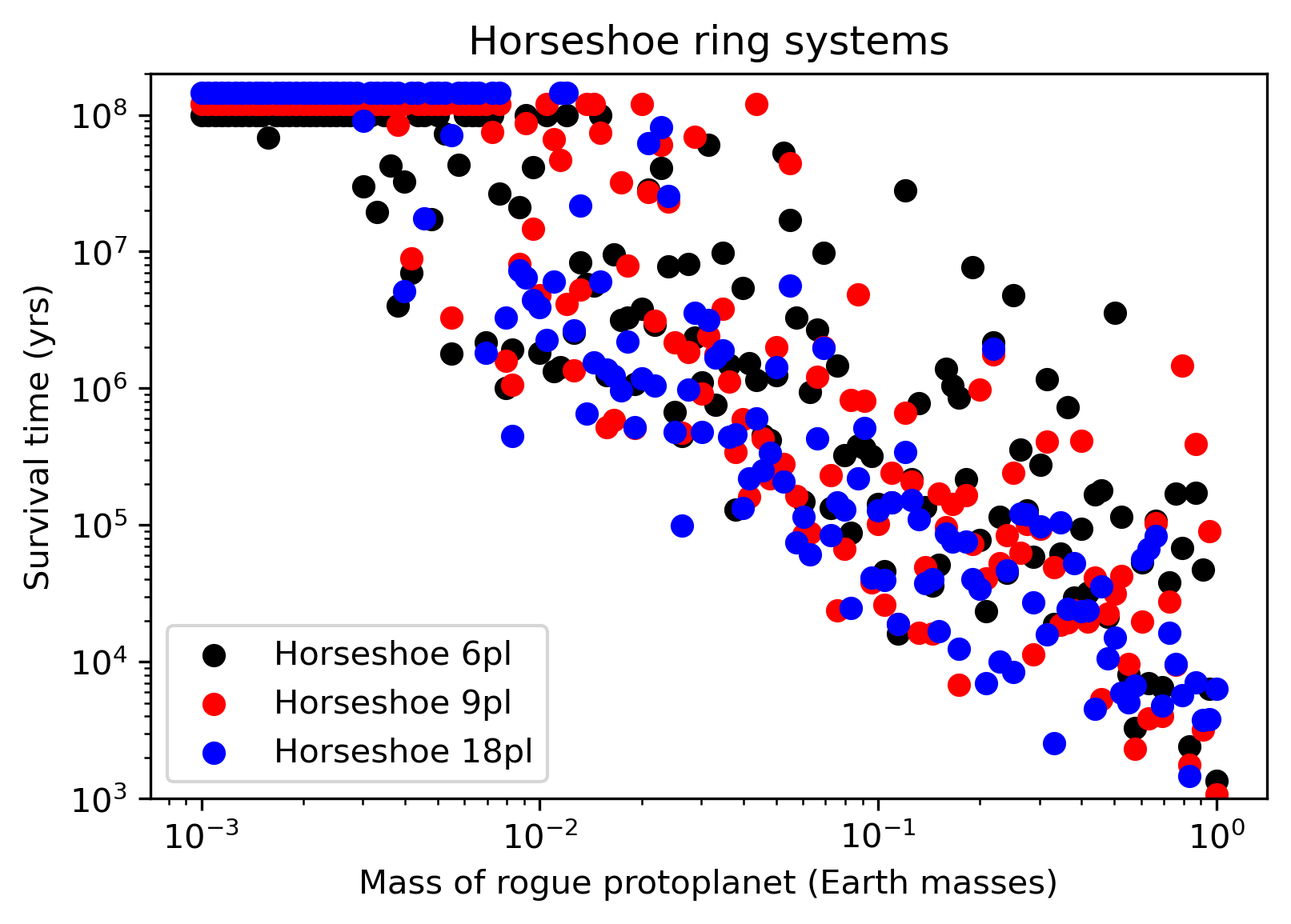}
    \caption{Survival of different co-orbital systems as a function of the mass of the rogue planetary embryo that was introduced into the system.  The colors correspond to co-orbital systems with different number of planets.  Among simulations that survived for the full 100 Myr integration, we introduced small vertical shifts for visibility.}
    \label{fig:survival}
\end{figure*}

In this paper we numerically assess the dynamical fragility of rings of co-orbital planets.  We invoke the existence of a single, rogue protoplanet that dynamically perturbs the systems. We treat the mass of the rogue protoplanet as a free parameter, and run simulations with different numbers of co-orbital planets (6, 9, 18 or 42) in both stationary and horseshoe configurations.

\section{Numerical Experiment}

Our general setup is analogous to that of \cite{raymond22}, which assessed the perturbations that would disrupt the multi-resonant orbital structure of the TRAPPIST-1 exoplanet system~\citep{gillon17,luger17,agol21}.  The main difference is that in this study we only consider the perturbations from a single rogue protoplanet and neglect the case of a swarm of rogue planetesimals.

\subsection{Simulations}

Our simulations started with two components: a ring of co-orbital planets and a single rogue protoplanet.  We tested seven different co-orbital rings: three horseshoe constellations and four stationary rings.  The horseshoe rings contained 6, 9, and 18 Earth-mass planets each. We chose rings that were long-term stable from our previous study of such systems, by simply starting the planets spread far enough apart, with initial separations of 25 mutual Hill radii~\citep[see Fig.~1 in ][]{raymond23}.  The stationary planet rings contained 6, 9, 18 and 42 planets of $1 \mearth$~each, and were constructed by simply spreading the planets evenly in mean anomaly along a near-circular (eccentricity of $10^{-5}$), co-planar orbit at 1 au. We first verified that each of the seven co-orbital rings was long-term stable by running 10 Gyr simulations using the hybrid algorithm in the {\tt Mercury} integration package~\citep{chambers99}.  

The orbit of the rogue protoplanet was chosen to cross the ring's orbital radius.  The protoplanet's perihelion distance was randomly chosen between 0.5 and 1 au, its semimajor axis between 1.5 and 2.5 au, and its inclination between zero and $10^\circ$.  While the exact orbital distribution of the rogue protoplanets does affect their angular momentum, it has only a small effect on the stability of the perturbed system~\citep{raymond22}.  We therefore did not systematically vary the rogue protoplanet's orbit.  

For each ring of co-orbital planets we ran 150 simulations varying the mass of the rogue protoplanet in the range of $0.001-1 \mearth$, sampled logarithmically.  This mass range was chosen after a few test simulations. Each simulation was integrated for 100 million years or until a collision or ejection occurred, again using the hybrid integrator in the {\tt Mercury} integration package~\citep{chambers99}, which we showed in~\cite{raymond23} to be as reliable as the Bulirsch-Stoer method (with acccuracy parameter of $10^{-15}$; see Appendix A in that paper). For each system that remained stable for 100 million years, we then ran a short simulation (in most cases of $1000$~years) with high-frequency outputs to assess the dynamical configuration of the final, perturbed system.

\subsection{Results}

Figure~\ref{fig:survival} shows how long each system survived as a function of the mass of the rogue protoplanet, for both stationary (left) and horseshoe (right) systems.  This Figure makes three points.  First, the critical mass above which a rogue protoplanet disrupts rings of co-orbital planets is close to a lunar mass ($\sim 0.01 \mearth$).  Second, the number of planets in a given co-orbital ring has little to no effect on the ring's stability.  Stationary ring systems with 6, 9, 18, and 42 planets remained stable with rogue protoplanet masses of $0.0347$, $0.0191$, $0.0263$, and $0.0263 \mearth$, respectively.  Among horseshoe constellation systems with 6, 9, and 18 planets the maximum stable rogue protoplanet masses were $0.0151$, $0.0437$, and $0.0120 \mearth$, respectively.  Finally, stationary ring systems are more resistant to perturbations than horseshoe systems.  While the maximum rogue protoplanet mass that allowed for stability was similar between the horseshoe and stationary ring systems, more than three times more simulations with $\mathrm{M_{Rogue}} > 0.01\mearth$ remained stable in stationary systems than in horseshoe systems, even if only considering the cases with 6, 9, or 18 planets.  

\begin{figure}
	\includegraphics[width=\columnwidth]{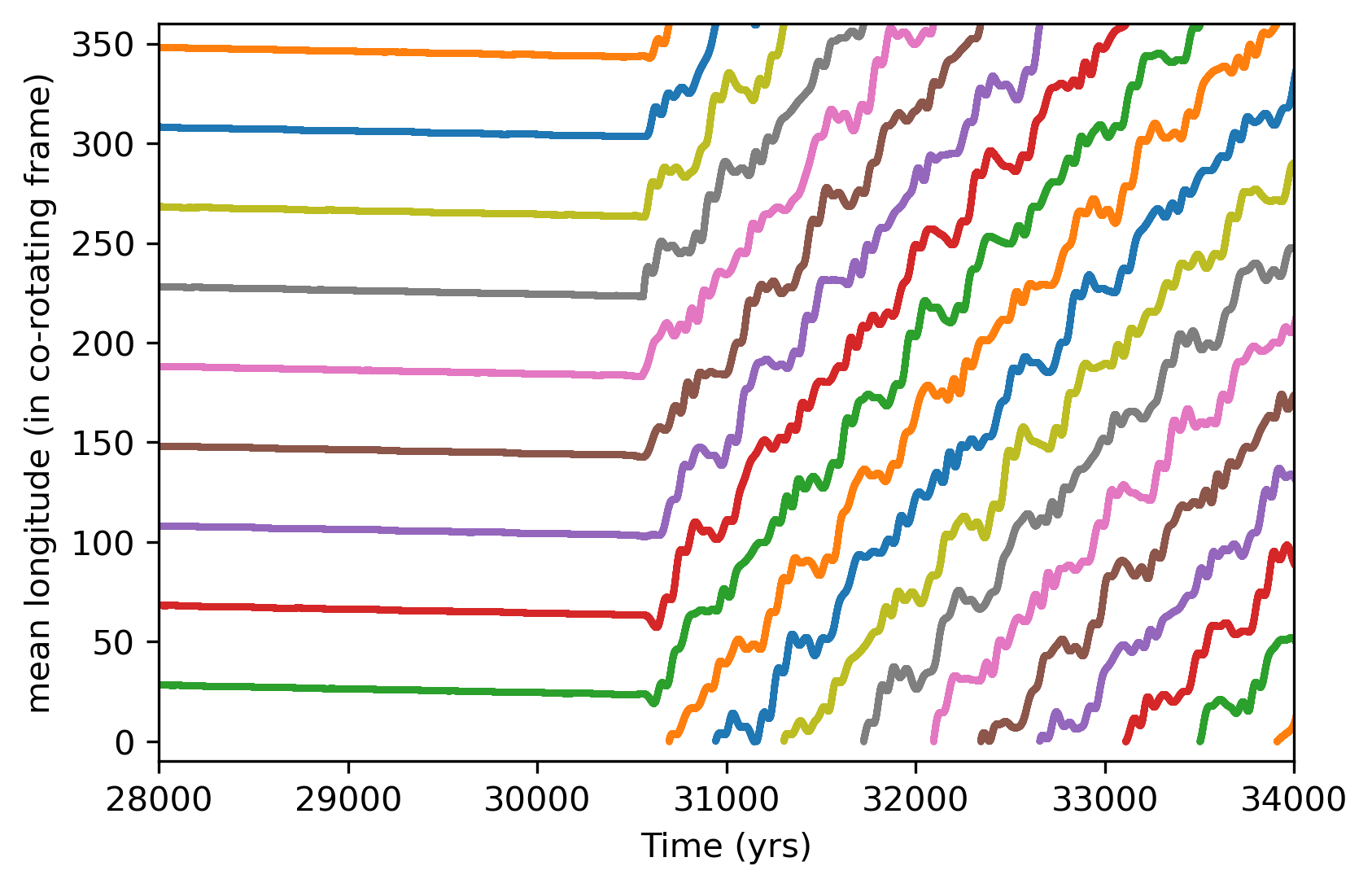}
\caption{Transformation of a stationary co-orbital ring into a horseshoe constellation system. The mean longitude of each planets is shown relative to the mean mean motion of a planet at 1 au (moving at $360^\circ \, \mathrm{yr^{-1}}$). The change in dynamical state was triggered by a close encounter between the rogue protoplanet and a planet at $t \sim 30,500$~years. }
    \label{fig:transition}
\end{figure}

After being perturbed, the dynamics of horseshoe constellations does not change appreciably (at least, among those that remain stable). However, the dynamics of stationary rings can change significantly.  Figure~\ref{fig:transition} shows a 9-planet stationary ring being perturbed by a lunar-mass ($0.12 \mearth$) protoplanet.  Before any significant close encounters take place, the planets remain evenly spread out in longitude.  However, after a strong close encounter between the rogue protoplanet and one planet in the ring after 30,500 years, the system transitions into a horseshoe constellation state.  The planets all precess much faster than in the stationary configuration, and undergo periodic mutual encounters that cause horseshoe librations (visible as `wiggles' in Fig.~\ref{fig:transition}).  This change in orbital configuration can also be seen in the planets' orbital semimajor axes.  The semimajor axes of planets in stationary rings remain perfectly constant in time.  However, each horseshoe oscillation causes an exchange in orbital angular momentum such that the semimajor axes of planets in the horseshoe regime are not fixed in time but rather oscillate. The dynamics of horseshoe oscillations, and the transition between the tadpole and horseshoe regimes in the restricted case with two planets, depend largely on the Jacobi constant~\citep[see][]{dermott81a,murraydermott99}. In our simulations we find a gradient of outcomes between stationary rings and horseshoe systems (see below), implying that in the case of many planets there is no sharp transition between the two regimes. 

\begin{figure}
 	\includegraphics[width=0.49\columnwidth]{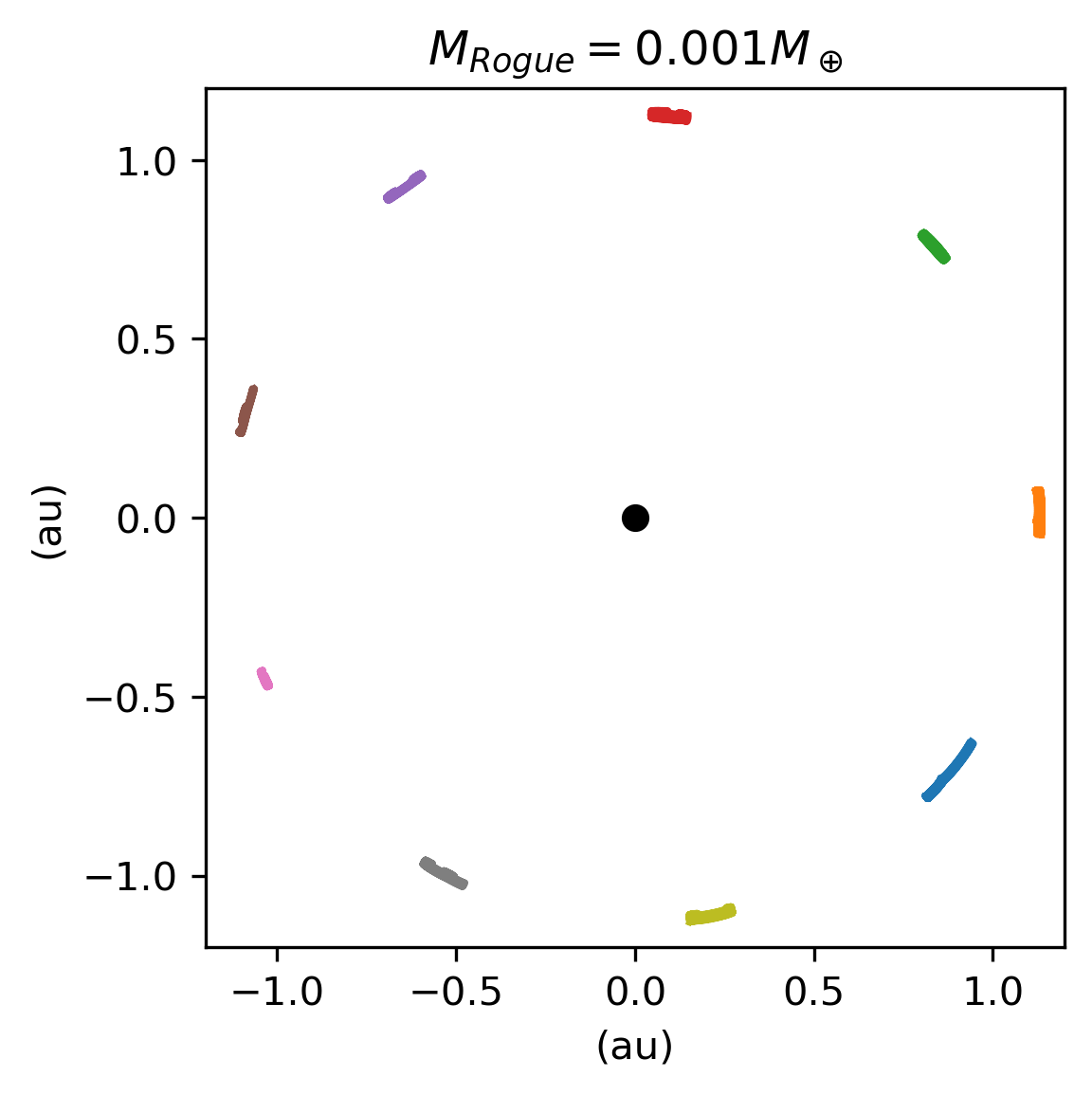} 
  	\includegraphics[width=0.49\columnwidth]{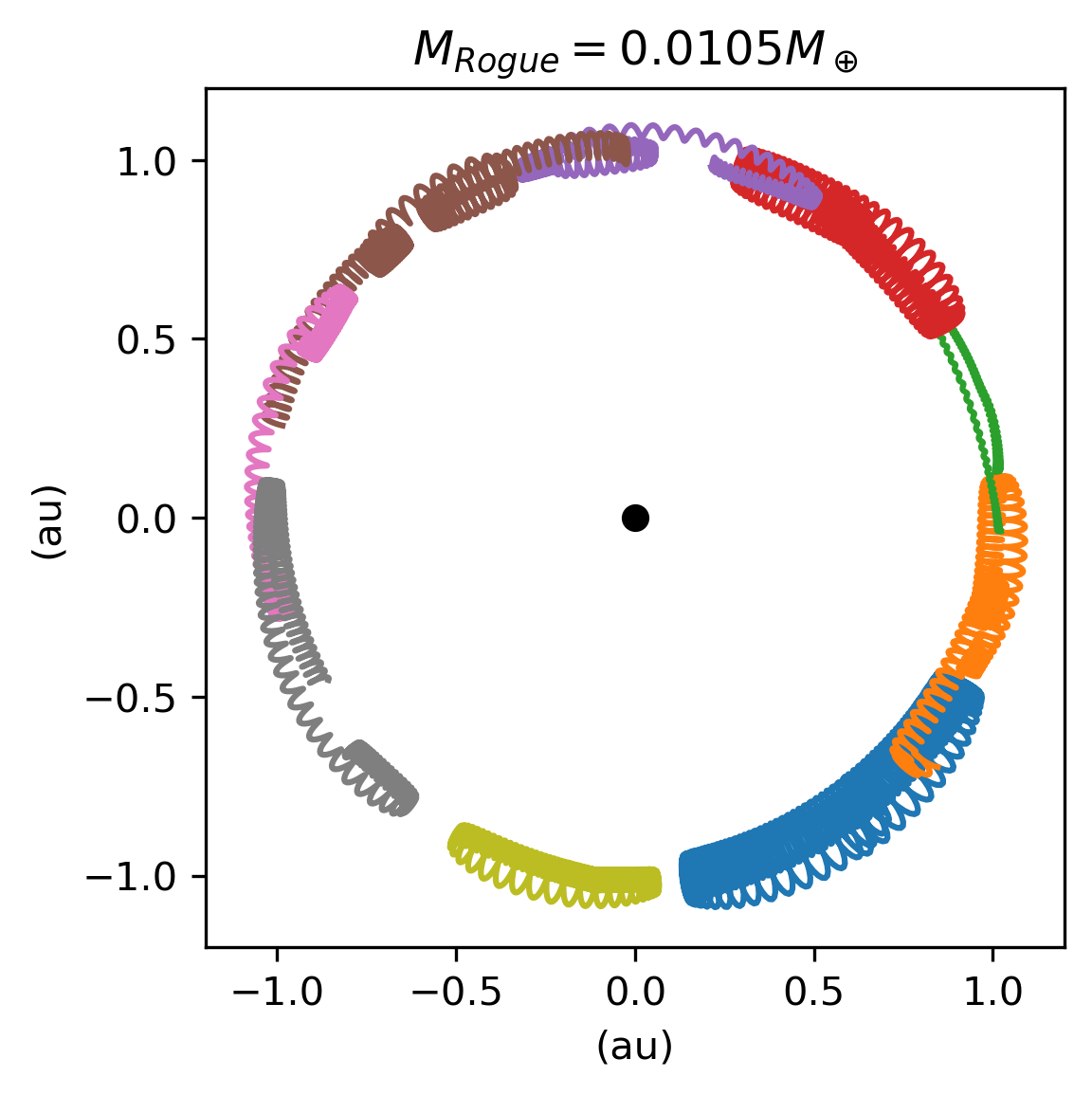}
        \caption{Two 9-planet ring systems that survived for 100 Myr but ended up in different dynamical states. Each panel shows the positions of each of the planets in the x-y plane of their orbit over a $\sim$100-year time span as viewed in a frame that is co-moving at the mean motion of an isolated planet at 1 au. The radial excursion of the planets has been enhanced by a factor of ten. The simulation on the left was barely perturbed by its $0.001 \mearth$ rogue embryo and remains in a near-stationary state.  In contrast, the simulation on the right was perturbed by its roughly Moon-mass ($0.0105 \mearth$) embryo into a horseshoe constellation.
        }
    \label{fig:examples}
\end{figure}

Figure~\ref{fig:examples} shows the short-term (100-yr) relative movement of co-orbital rings in two systems that both started off as stationary 9-planet co-orbital rings. The mass of the rogue protoplanet varied by an order of magnitude between the two systems, with $\mathrm{M_{Rogue}} = 0.001 \mearth$ (left panel) and $0.0105 \mearth$ (right panel).  The system with $\mathrm{M_{Rogue}} = 0.001 \mearth$ was only weakly-perturbed: the planets' radial excursions (and semimajor axis oscillations) remained minimal and the system maintained its stationary configuration. This is the same behavior seen in the system from Fig.~\ref{fig:transition} before the rogue protoplanet close encounter at 30,500 years.  In contrast, the system with $\mathrm{M_{Rogue}} = 0.0105 \mearth$ was perturbed into a horseshoe constellation system, with significant semimajor axis oscillations and frequent encounters between planets.  This is similar to the behavior of the system from Fig.~\ref{fig:transition} after the rogue protoplanet close encounter.  

\begin{figure*}
	\includegraphics[width=\columnwidth]{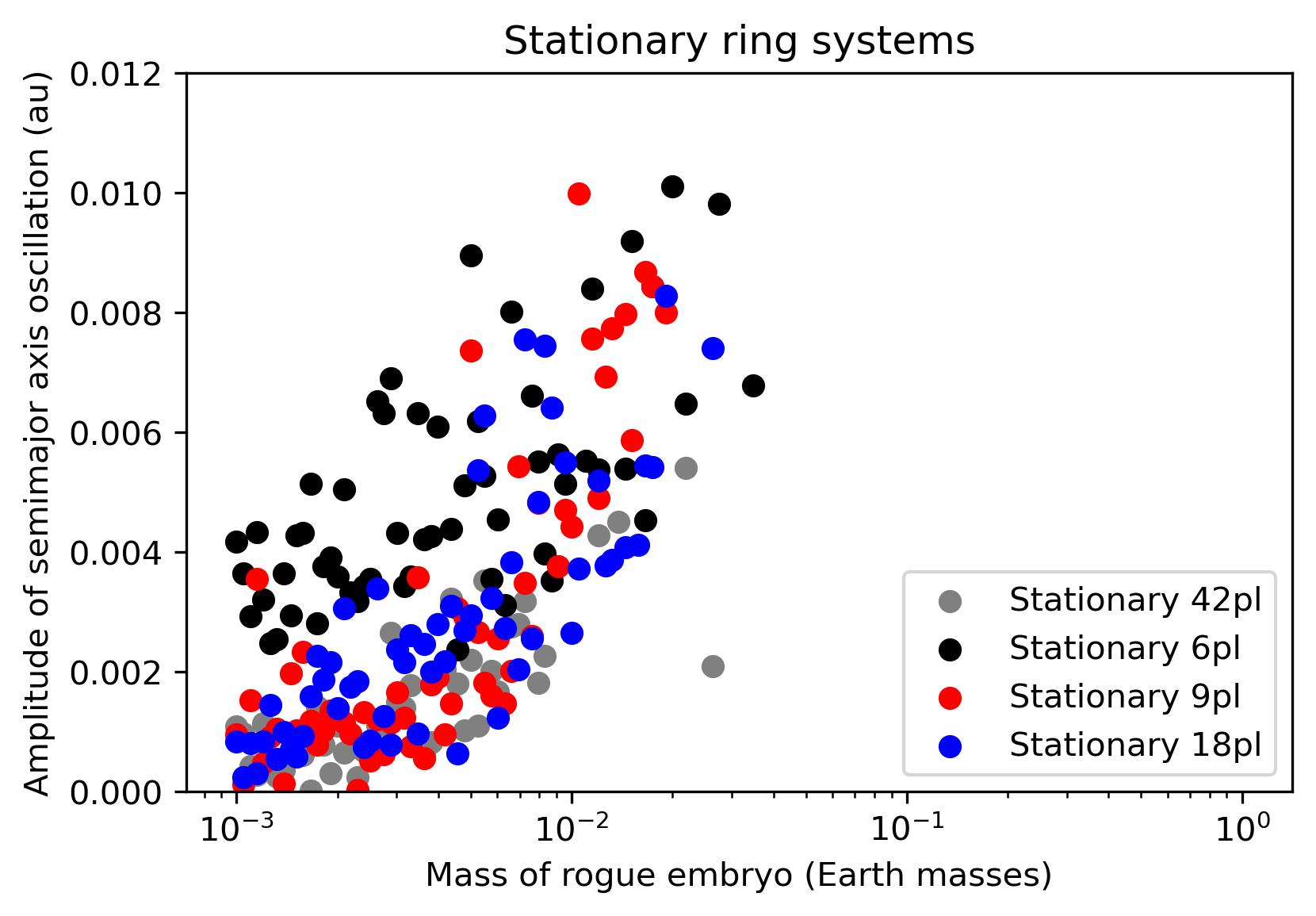}
 	\includegraphics[width=\columnwidth]{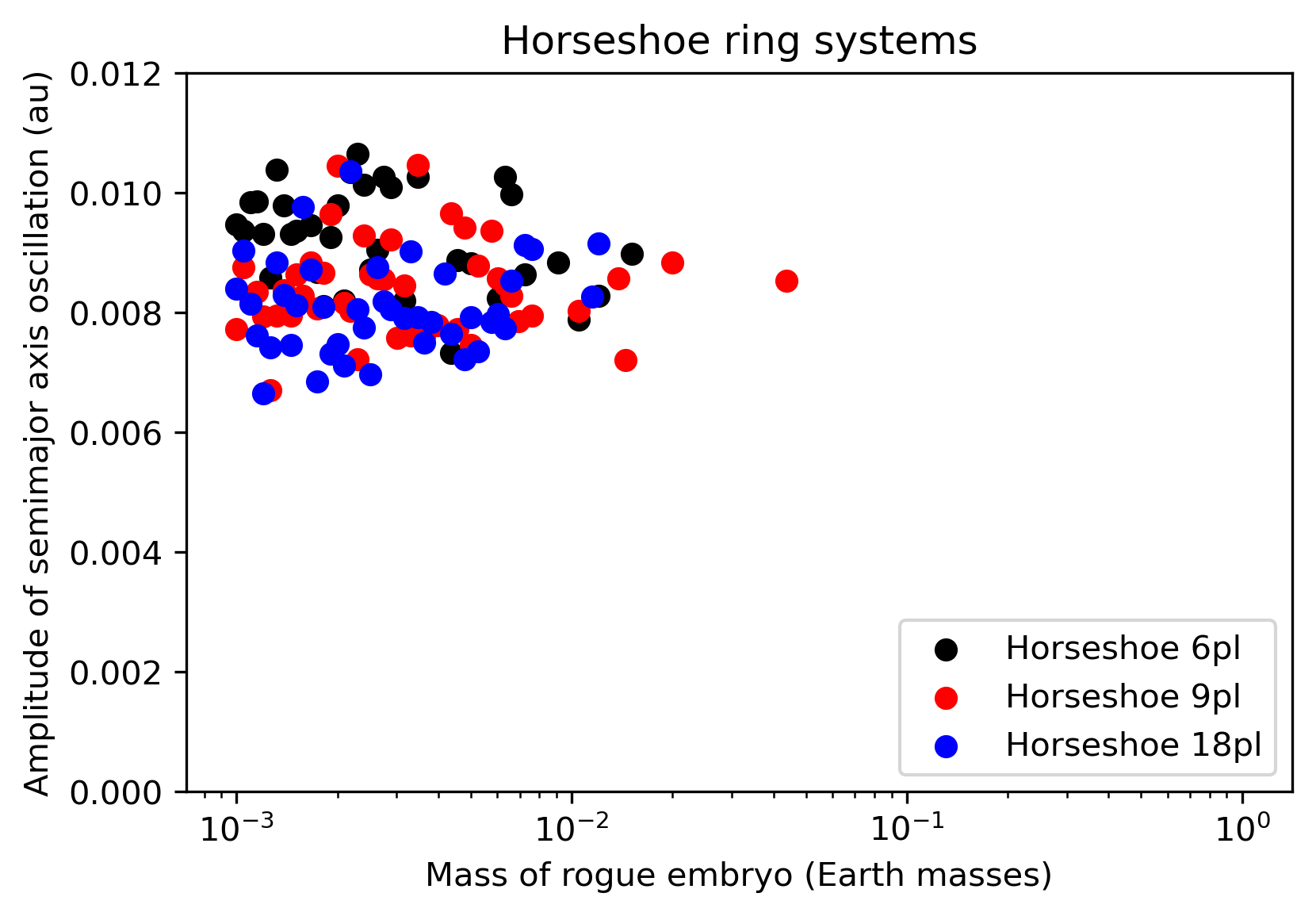}
    \caption{Amplitude of semimajor axis oscillations among systems that were stable for 100 Myr.  }
    \label{fig:sma_osc}
\end{figure*}

Figure~\ref{fig:sma_osc} shows the oscillation amplitudes of the semimajor axes of planets in each surviving co-orbital system as a function of the rogue protoplanet mass.  The horseshoe systems all have similar amplitudes of oscillation of a little less than 0.01~au, regardless of $\mathrm{M_{Rogue}}$. In contrast, for stationary ring systems the semimajor axis oscillation amplitude is a strong function of $\mathrm{M_{Rogue}}$.  This is because the strength of encounters with the rogue protoplanet determine the system's orbital state.  There is a continuum of semimajor axis oscillation amplitudes (and frequencies) governed by the closest approaches between planets~\citep[and the Jacobi constant; see][]{dermott81a}. This can be explained naturally if the amplitude of semimajor axis oscillation increases for more massive rogue protoplanets.  

The fact that no surviving systems have semimajor axis oscillation amplitudes larger than $\sim 0.01$~au is a clue that the patterns from Fig.~\ref{fig:sma_osc} are the result of survivorship bias: any system that was excited to a higher level of oscillation was destabilized.  This makes sense from a dynamical point of view, as higher oscillation amplitudes correlate with closer horseshoe encounters~\citep{dermott81a}, yet encounters closer than $\sim 5$ Hill radii (or $\sim 4$ mutual Hill radii for near-equal planet masses) result in disruption of the horseshoe system~\citep{cuk12b}.  Rogue protoplanets with masses larger than a few lunar masses excite unstable horseshoe librations, leading to destabilization on a timescale given by the protoplanet mass (see Fig.~\ref{fig:survival}). Horseshoe systems are more fragile than stationary ring systems simply because stationary ring systems start with zero semimajor axis oscillation amplitudes whereas horseshoes start with significant amplitudes.  In both types of systems, these oscillation amplitudes are amplified by perturbations from the rogue protoplanet. Even though individual perturbation events do not strictly always increase the oscillation amplitude (because this depends on the exact geometry of close encounters), horseshoe systems start out closer to the stability limit than stationary ring systems.  Indeed, these inherent semimajor axis oscillations -- which are a result of the construction of horseshoe constellation systems -- are close enough to the stability limit that, although the probability of stability is a strong function of $\mathrm{M_{Rogue}}$, the oscillation amplitude is not.

\begin{figure}
	\includegraphics[width=\columnwidth]{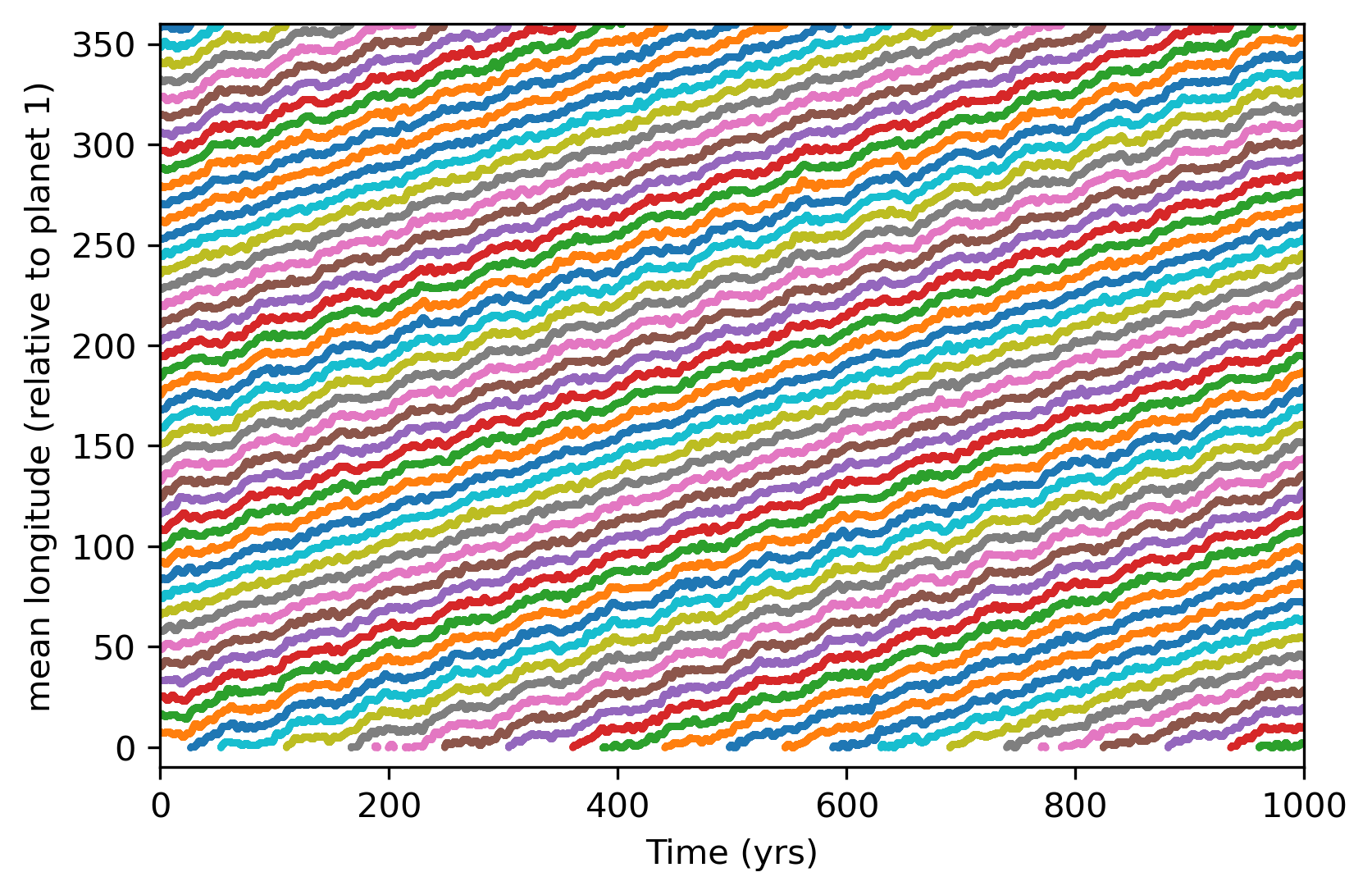}
\caption{Evolution of the most strongly-perturbed surviving 42-planet ring from our simulations. In this case, the rogue protoplanet mass was $\mathrm{M_{Rogue}} = 0.0219 \mearth$. As in Fig.~\ref{fig:transition}, the mean longitude of each planets is shown relative to the mean mean motion of a planet at 1 au (moving at $360^\circ \, \mathrm{yr^{-1}}$). Despite the repeating colors, each continuous curve corresponds to a single planet.}
    \label{fig:42}
\end{figure}

The 42-Earth system is maximally-packed for long-term stable stationary rings~\citep{salo88,smith10}. In the co-orbital ring systems presented in Section 2 -- with 6, 9 or 18 planets -- there exist stable configurations in both the realm of stationary and horseshoe systems. In the 42-planet case, there is no corresponding stable horseshoe constellation system, as those have only been shown to exist with up to 24 planets~\citep{raymond23}. 

The 42-Earth stationary ring systems behave in a very similar way to stationary rings with fewer planets.  The 42-planet systems are modestly less stable, with instability times that tend to be shorter for a given rogue protoplanet mass (Fig.~\ref{fig:survival}), although the maximum stable rogue protoplanet mass for stability is roughly the same as in other systems.  

There are no surviving 42-Earth stationary ring systems with large-amplitude semimajor axis oscillations.  The maximum oscillation amplitude for 42-Earth rings is less than 0.006 au, significantly lower than for 6-, 9-, or 18-Earth rings. This is likely a consequence of the instability of horseshoe constellation systems above a given number of planets. Figure~\ref{fig:42} shows the evolution of the most strongly-perturbed 42-Earth stationary ring that remained stable.  While the mean longitudes of the planets do not evolve in a completely smooth way, each planet moves in concert such that there are no between planets closer than 10 mutual Hill radii.  While the planets' semimajor axes do oscillate, this appears to be due to cumulative relatively distant perturbations rather than horseshoe-type encounters.

\section{Discussion}

\subsection{Implications for the detectability of co-orbital rings}

In previous work we have put forth the idea that a highly-advanced civilization could construct a planetary system to act as a cosmic signpost of its presence. We called such systems `SETI beacons'.  In \cite{clement22} we showed how SETI beacons could be made from multi-resonant planetary systems in which the period ratios of neighboring planets produced a sequence of integers that would be recognizable as being non-natural (such as consecutive prime numbers).  In \cite{raymond23} we proposed that horseshoe constellation systems with many planets may represent potential SETI beacons.  

Stationary co-orbital rings of planets represent another flavor of SETI beacon.  Yet, given the results of this study, we might expect horseshoe constellation systems to be more common than stationary rings.  The reason is simply that there is nowhere in the Universe that is immune from gravitational perturbations.  If one could construct a finely-tuned stationary co-orbital system in isolation, it would invariably be affected by other nearby objects.  It may be advantageous to build such a system late in a star's lifetime, as the likelihood of perturbation would decrease in time as leftover planetesimals were dynamically and collisionally removed~\citep[and knowing that co-orbital rings can survive post-main sequence evolution;][]{raymond23}.  Perhaps the most protected cosmic environment within reach of detection would be relatively close to a star on the outskirts of a galaxy~\citep[not too close or star-planet tidal effects might compromise dynamical stability;][]{rodriguez13,leleu15}. Yet the whole point of a SETI beacon is to be long-lasting, and even exceedingly rare perturbations will occur given sufficient time (and this applies regardless of the exact orbital configuration of the system).  For instance, our Solar System will likely undergo a strong dynamical instability triggered by the close flyby of a star after the Sun has become a white dwarf~\citep{zink20}.  A carefully-selected galactic environment could reduce the chances or the magnitude of external perturbations~\citep[although the Galaxy's stellar environment will change drastically after the merger between the Milky Way and M31; e.g.,][]{cox08}.  Yet even relatively modest perturbations (akin to the gravitational influence of a Moon-sized rogue protoplanet) would transform a stationary ring system into a horseshoe constellation.

\subsection{Limitations}
This study was admittedly simplified and limited.  The parameter space of co-orbital planets is vast~\citep[see, for example,][for a discussion]{laughlin02,giuppone12}, and our study only covers a narrow region.  Given the relatively simple, proof-of-concept nature of this work we did not think that there was much to gain by testing other system parameters such as the planet mass and orbital radius or the rogue protoplanet's mass and orbital characteristics.  This choice means that we limited ourselves to relatively short, intense perturbations and did not consider slower, more continuous ones such as those generated by a population of rogue planetesimals~\citep[see][]{raymond22}, other planets orbiting the same star, or even Galactic tides (for wide-orbit co-orbital rings).  Tidal dissipation may also be important in co-orbital systems, especially on long timescales or in systems that are relatively close to their host stars~\citep{rodriguez13,leleu15}.  In addition, cross-tides", whereby one planet is torqued by the tidal bulge generated on the star by another planet~\citep{touma94,nerondesurgy97,lainey17}, may also be worth taking into account in future studies.

Finally, we emphasize that even though co-orbital rings of planets may be extremely rare, it would be a monumental event to discover such a system.  They may be within the reach of current and upcoming exoplanet surveys such as TESS and PLATO~\citep[e.g.][]{barclay18}, and we strongly encourage observers to keep them in mind when analyzing unexpected signals.

\section*{Acknowledgements}
We thank the anonymous referee for a helpful report that improved the paper. S.N.R. thanks the CNRS's PNP and MITI programs for support.  V.S.M. and S.N.R. also acknowledge support from NASA’s NExSS Virtual Planetary Laboratory, funded under NASA Astrobiology Program grant 80NSSC18K0829. A.I. acknowledges support from NASA grant 80NSSC18K0828 and the Welch Foundation grant No. C-2035-20200401.

%%%%%%%%%%%%%%%%%%%%%%%%%%%%%%%%%%%%%%%%%%%%%%%%%%
%\section*{Animations}

%%%%%%%%%%%%%%%%%%%%%%%%%%%%%%%%%%%%%%%%%%%%%%%%%%
\section*{Data Availability}

All simulations and analysis code in this paper will be made available upon reasonable request.

\bsp	% typesetting comment
\label{lastpage}

\begin{thebibliography}{}
\makeatletter
\relax
\def\mn@urlcharsother{\let\do\@makeother \do\$\do\&\do\#\do\^\do\_\do\%\do\~}
\def\mn@doi{\begingroup\mn@urlcharsother \@ifnextchar [ {\mn@doi@}
  {\mn@doi@[]}}
\def\mn@doi@[#1]#2{\def\@tempa{#1}\ifx\@tempa\@empty \href
  {http://dx.doi.org/#2} {doi:#2}\else \href {http://dx.doi.org/#2} {#1}\fi
  \endgroup}
\def\mn@eprint#1#2{\mn@eprint@#1:#2::\@nil}
\def\mn@eprint@arXiv#1{\href {http://arxiv.org/abs/#1} {{\tt arXiv:#1}}}
\def\mn@eprint@dblp#1{\href {http://dblp.uni-trier.de/rec/bibtex/#1.xml}
  {dblp:#1}}
\def\mn@eprint@#1:#2:#3:#4\@nil{\def\@tempa {#1}\def\@tempb {#2}\def\@tempc
  {#3}\ifx \@tempc \@empty \let \@tempc \@tempb \let \@tempb \@tempa \fi \ifx
  \@tempb \@empty \def\@tempb {arXiv}\fi \@ifundefined
  {mn@eprint@\@tempb}{\@tempb:\@tempc}{\expandafter \expandafter \csname
  mn@eprint@\@tempb\endcsname \expandafter{\@tempc}}}

\bibitem[\protect\citeauthoryear{{Agol} et~al.,}{{Agol} et~al.}{2021}]{agol21}
{Agol} E.,  et~al., 2021, \mn@doi [The Planetary Science Journal]
  {10.3847/PSJ/abd022}, \href
  {https://ui.adsabs.harvard.edu/abs/2021PSJ.....2....1A} {2, 1}

\bibitem[\protect\citeauthoryear{{Barclay}, {Pepper}  \& {Quintana}}{{Barclay}
  et~al.}{2018}]{barclay18}
{Barclay} T.,  {Pepper} J.,   {Quintana} E.~V.,  2018, \mn@doi [\apjs]
  {10.3847/1538-4365/aae3e9}, \href
  {https://ui.adsabs.harvard.edu/abs/2018ApJS..239....2B} {239, 2}

\bibitem[\protect\citeauthoryear{{Chambers}}{{Chambers}}{1999}]{chambers99}
{Chambers} J.~E.,  1999, \mn@doi [\mnras] {10.1046/j.1365-8711.1999.02379.x},
  \href {http://adsabs.harvard.edu/abs/1999MNRAS.304..793C} {304, 793}

\bibitem[\protect\citeauthoryear{{Clement}, {Raymond}, {Veras}  \&
  {Kipping}}{{Clement} et~al.}{2022}]{clement22}
{Clement} M.~S.,  {Raymond} S.~N.,  {Veras} D.,   {Kipping} D.,  2022, \mn@doi
  [\mnras] {10.1093/mnras/stac1234}, \href
  {https://ui.adsabs.harvard.edu/abs/2022MNRAS.513.4945C} {513, 4945}

\bibitem[\protect\citeauthoryear{{Cox} \& {Loeb}}{{Cox} \&
  {Loeb}}{2008}]{cox08}
{Cox} T.~J.,  {Loeb} A.,  2008, \mn@doi [\mnras]
  {10.1111/j.1365-2966.2008.13048.x}, \href
  {https://ui.adsabs.harvard.edu/abs/2008MNRAS.386..461C} {386, 461}

\bibitem[\protect\citeauthoryear{{Cresswell} \& {Nelson}}{{Cresswell} \&
  {Nelson}}{2009}]{cresswell09}
{Cresswell} P.,  {Nelson} R.~P.,  2009, \mn@doi [\aap]
  {10.1051/0004-6361:200810705}, \href
  {http://adsabs.harvard.edu/abs/2009A%26A...493.1141C} {493, 1141}

\bibitem[\protect\citeauthoryear{{{\'C}uk}, {Hamilton}  \& {Holman}}{{{\'C}uk}
  et~al.}{2012}]{cuk12b}
{{\'C}uk} M.,  {Hamilton} D.~P.,   {Holman} M.~J.,  2012, \mn@doi [\mnras]
  {10.1111/j.1365-2966.2012.21964.x}, \href
  {https://ui.adsabs.harvard.edu/abs/2012MNRAS.426.3051C} {426, 3051}

\bibitem[\protect\citeauthoryear{{Dermott} \& {Murray}}{{Dermott} \&
  {Murray}}{1981a}]{dermott81a}
{Dermott} S.~F.,  {Murray} C.~D.,  1981a, \mn@doi [\icarus]
  {10.1016/0019-1035(81)90147-0}, \href
  {https://ui.adsabs.harvard.edu/abs/1981Icar...48....1D} {48, 1}

\bibitem[\protect\citeauthoryear{{Dermott} \& {Murray}}{{Dermott} \&
  {Murray}}{1981b}]{dermott81b}
{Dermott} S.~F.,  {Murray} C.~D.,  1981b, \mn@doi [\icarus]
  {10.1016/0019-1035(81)90148-2}, \href
  {https://ui.adsabs.harvard.edu/abs/1981Icar...48...12D} {48, 12}

\bibitem[\protect\citeauthoryear{{Gillon} et~al.,}{{Gillon}
  et~al.}{2017}]{gillon17}
{Gillon} M.,  et~al., 2017, \mn@doi [\nat] {10.1038/nature21360}, \href
  {http://adsabs.harvard.edu/abs/2017Natur.542..456G} {542, 456}

\bibitem[\protect\citeauthoryear{{Giuppone}, {Ben{\'\i}tez-Llambay}  \&
  {Beaug{\'e}}}{{Giuppone} et~al.}{2012}]{giuppone12}
{Giuppone} C.~A.,  {Ben{\'\i}tez-Llambay} P.,   {Beaug{\'e}} C.,  2012, \mn@doi
  [\mnras] {10.1111/j.1365-2966.2011.20310.x}, \href
  {https://ui.adsabs.harvard.edu/abs/2012MNRAS.421..356G} {421, 356}

\bibitem[\protect\citeauthoryear{{Izidoro}, {Ogihara}, {Raymond}, {Morbidelli},
  {Pierens}, {Bitsch}, {Cossou}  \& {Hersant}}{{Izidoro}
  et~al.}{2017}]{izidoro17}
{Izidoro} A.,  {Ogihara} M.,  {Raymond} S.~N.,  {Morbidelli} A.,  {Pierens} A.,
   {Bitsch} B.,  {Cossou} C.,   {Hersant} F.,  2017, \mn@doi [\mnras]
  {10.1093/mnras/stx1232}, \href
  {http://adsabs.harvard.edu/abs/2017MNRAS.470.1750I} {470, 1750}

\bibitem[\protect\citeauthoryear{{Lainey} et~al.,}{{Lainey}
  et~al.}{2017}]{lainey17}
{Lainey} V.,  et~al., 2017, \mn@doi [\icarus] {10.1016/j.icarus.2016.07.014},
  \href {https://ui.adsabs.harvard.edu/abs/2017Icar..281..286L} {281, 286}

\bibitem[\protect\citeauthoryear{{Laughlin} \& {Chambers}}{{Laughlin} \&
  {Chambers}}{2002}]{laughlin02}
{Laughlin} G.,  {Chambers} J.~E.,  2002, \mn@doi [\aj] {10.1086/341173}, \href
  {https://ui.adsabs.harvard.edu/abs/2002AJ....124..592L} {124, 592}

\bibitem[\protect\citeauthoryear{{Leleu}, {Robutel}  \& {Correia}}{{Leleu}
  et~al.}{2015}]{leleu15}
{Leleu} A.,  {Robutel} P.,   {Correia} A. C.~M.,  2015, \mn@doi [\aap]
  {10.1051/0004-6361/201526175}, \href
  {https://ui.adsabs.harvard.edu/abs/2015A&A...581A.128L} {581, A128}

\bibitem[\protect\citeauthoryear{{Lillo-Box}, {Barrado}, {Figueira}, {Leleu},
  {Santos}, {Correia}, {Robutel}  \& {Faria}}{{Lillo-Box}
  et~al.}{2018a}]{lillobox18a}
{Lillo-Box} J.,  {Barrado} D.,  {Figueira} P.,  {Leleu} A.,  {Santos} N.~C.,
  {Correia} A.~C.~M.,  {Robutel} P.,   {Faria} J.~P.,  2018a, \mn@doi [\aap]
  {10.1051/0004-6361/201730652}, \href
  {https://ui.adsabs.harvard.edu/abs/2018A&A...609A..96L} {609, A96}

\bibitem[\protect\citeauthoryear{{Lillo-Box} et~al.,}{{Lillo-Box}
  et~al.}{2018b}]{lillobox18b}
{Lillo-Box} J.,  et~al., 2018b, \mn@doi [\aap] {10.1051/0004-6361/201833312},
  \href {https://ui.adsabs.harvard.edu/abs/2018A&A...618A..42L} {618, A42}

\bibitem[\protect\citeauthoryear{{Luger} et~al.,}{{Luger}
  et~al.}{2017}]{luger17}
{Luger} R.,  et~al., 2017, \mn@doi [Nature Astronomy]
  {10.1038/s41550-017-0129}, \href
  {http://adsabs.harvard.edu/abs/2017NatAs...1E.129L} {1, 0129}

\bibitem[\protect\citeauthoryear{{Murray} \& {Dermott}}{{Murray} \&
  {Dermott}}{1999}]{murraydermott99}
{Murray} C.~D.,  {Dermott} S.~F.,  1999, {Solar system dynamics}.
Cambridge university press

\bibitem[\protect\citeauthoryear{{Neron de Surgy} \& {Laskar}}{{Neron de Surgy}
  \& {Laskar}}{1997}]{nerondesurgy97}
{Neron de Surgy} O.,  {Laskar} J.,  1997, \aap, \href
  {https://ui.adsabs.harvard.edu/abs/1997A&A...318..975N} {318, 975}

\bibitem[\protect\citeauthoryear{{Raymond}, {Boulet}, {Izidoro}, {Esteves}  \&
  {Bitsch}}{{Raymond} et~al.}{2018}]{raymond18b}
{Raymond} S.~N.,  {Boulet} T.,  {Izidoro} A.,  {Esteves} L.,   {Bitsch} B.,
  2018, \mn@doi [\mnras] {10.1093/mnrasl/sly100}, \href
  {http://adsabs.harvard.edu/abs/2018MNRAS.479L..81R} {479, L81}

\bibitem[\protect\citeauthoryear{{Raymond} et~al.,}{{Raymond}
  et~al.}{2022}]{raymond22}
{Raymond} S.~N.,  et~al., 2022, \mn@doi [Nature Astronomy]
  {10.1038/s41550-021-01518-6}, \href
  {https://ui.adsabs.harvard.edu/abs/2022NatAs...6...80R} {6, 80}

\bibitem[\protect\citeauthoryear{{Raymond}, {Veras}, {Clement}, {Izidoro},
  {Kipping}  \& {Meadows}}{{Raymond} et~al.}{2023}]{raymond23}
{Raymond} S.~N.,  {Veras} D.,  {Clement} M.~S.,  {Izidoro} A.,  {Kipping} D.,
  {Meadows} V.,  2023, \mn@doi [\mnras] {10.1093/mnras/stad643}, \href
  {https://ui.adsabs.harvard.edu/abs/2023MNRAS.521.2002R} {521, 2002}

\bibitem[\protect\citeauthoryear{{Rodr{\'\i}guez}, {Giuppone}  \&
  {Michtchenko}}{{Rodr{\'\i}guez} et~al.}{2013}]{rodriguez13}
{Rodr{\'\i}guez} A.,  {Giuppone} C.~A.,   {Michtchenko} T.~A.,  2013, \mn@doi
  [Celestial Mechanics and Dynamical Astronomy] {10.1007/s10569-013-9502-y},
  \href {https://ui.adsabs.harvard.edu/abs/2013CeMDA.117...59R} {117, 59}

\bibitem[\protect\citeauthoryear{{Rodr{\'\i}guez}, {Correa-Otto}  \&
  {Michtchenko}}{{Rodr{\'\i}guez} et~al.}{2019}]{rodriguez19}
{Rodr{\'\i}guez} A.,  {Correa-Otto} J.~A.,   {Michtchenko} T.~A.,  2019,
  \mn@doi [\mnras] {10.1093/mnras/stz1428}, \href
  {https://ui.adsabs.harvard.edu/abs/2019MNRAS.487.1973R} {487, 1973}

\bibitem[\protect\citeauthoryear{{Rowe} et~al.,}{{Rowe} et~al.}{2006}]{rowe06}
{Rowe} J.~F.,  et~al., 2006, \mn@doi [\apj] {10.1086/504252}, \href
  {https://ui.adsabs.harvard.edu/abs/2006ApJ...646.1241R} {646, 1241}

\bibitem[\protect\citeauthoryear{{Salo} \& {Yoder}}{{Salo} \&
  {Yoder}}{1988}]{salo88}
{Salo} H.,  {Yoder} C.~F.,  1988, \aap, \href
  {https://ui.adsabs.harvard.edu/abs/1988A&A...205..309S} {205, 309}

\bibitem[\protect\citeauthoryear{{Smith} \& {Lissauer}}{{Smith} \&
  {Lissauer}}{2010}]{smith10}
{Smith} A.~W.,  {Lissauer} J.~J.,  2010, \mn@doi [Celestial Mechanics and
  Dynamical Astronomy] {10.1007/s10569-010-9288-0}, \href
  {https://ui.adsabs.harvard.edu/abs/2010CeMDA.107..487S} {107, 487}

\bibitem[\protect\citeauthoryear{{Smith}, {Reitsema}, {Fountain}  \&
  {Larson}}{{Smith} et~al.}{1980}]{smith80}
{Smith} B.~A.,  {Reitsema} H.~J.,  {Fountain} J.~W.,   {Larson} S.~M.,  1980,
  in Bulletin of the American Astronomical Society. p.~727

\bibitem[\protect\citeauthoryear{{Touma} \& {Wisdom}}{{Touma} \&
  {Wisdom}}{1994}]{touma94}
{Touma} J.,  {Wisdom} J.,  1994, \mn@doi [\aj] {10.1086/117209}, \href
  {https://ui.adsabs.harvard.edu/abs/1994AJ....108.1943T} {108, 1943}

\bibitem[\protect\citeauthoryear{{Zink}, {Batygin}  \& {Adams}}{{Zink}
  et~al.}{2020}]{zink20}
{Zink} J.~K.,  {Batygin} K.,   {Adams} F.~C.,  2020, \mn@doi [\aj]
  {10.3847/1538-3881/abb8de}, \href
  {https://ui.adsabs.harvard.edu/abs/2020AJ....160..232Z} {160, 232}

\makeatother
\end{thebibliography}
\end{document}